\documentstyle[anta, twoside]{article}
\input{psfig.sty}

\def\Journal#1#2#3#4{(#1) {#2} {\bf #3}, #4}

\def\AAp{\em Astron. Astrophys.}

\def\ApJ{\em Astrophys.~J.}

\def\MNRAS{\em Mon. Not. R.~Astron. Soc.}

\def\PASP{\em Publ. Astron. Soc. Pac.}
\def\PASJ{\em Publ. Astron. Soc. Japan}
\def\SA{\em Sov. Astron.}
\newcommand{\HI}{{\rm H\,\scriptstyle I}}
\newcommand{\HII}{{\rm H\,\scriptstyle II}}
\begin{document}

\markboth{I. Yusifov \& \.I. K\"u\c c\"uk}{Galactic Distribution
          and the Luminosity Function of Pulsars}

\thispagestyle{plain}
\setcounter{page}{159}

\title{Galactic Distribution and the Luminosity Function of Pulsars}

\author{I. Yusifov and \.I. K\"u\c c\"uk}

\address{Department of Astronomy \& Space Sciences,
         Faculty of Arts \& Sciences,\\
              Erciyes University, Talas Yolu, 38039 Kayseri,
              Turkey\\}

\maketitle

\abstract{We have studied the radial distribution and luminosity
function of normal pulsars in the Galaxy, on the basis of the
ATNF Pulsar Catalogue where the distances are calculated
according to the new electron density model NE2001.\\
\noindent We show that the maximum of galactocentric distribution of
pulsars located near 3.5 kpc and the scale-length of this distribution
is $\sim 4$ kpc. The surface density of pulsars near the Galactic
center (GC) is equal or slightly higher than that in the solar
neighborhood. The local density and birth-rate of pulsars is 1.5 and 4
times higher than previous estimates, correspondingly before and after
beaming corrections. The dependence of these results on the NE2001
model and recommendations for further improvement of the electron
density distribution are discussed.\\
\noindent We derived the luminosity function of pulsars at frequencies
near 1400 MHz. In the limited band of luminosities, the luminosity
function has a slope near $-1$, but a new luminosity function may be
better fitted by a Log--Normal distribution.  }

\section{Introduction}

The high-frequency, sensitive Parkes Multibeam Pulsar Survey
(PMPS) (Manchester et al. 2001; Morris et al. 2002; Kramer et al. 2003)
revealed many more distant pulsars with high Dispersion Measures (DM).
These data may allow us to investigate in more detail the statistical
parameters and distribution of pulsars, especially in the central
regions of the Galaxy, which was almost impossible in previous
low-frequency and less-sensitive surveys. Furthermore, high-frequency
searches reveal many pulsars at the lower as well as at the high
luminosity ends of the luminosity function (LF) of pulsars. From the
evolutionary point of view it is of great interest for the
influence of such pulsars on the shape of pulsars' LF.

Considering the actuality of the problem and with the appearance of
publications with completely wrong interpretation of observational
data of pulsars (Guseinov et al. 2002a, 2002b), we decided to estimate
the density of pulsars at the GC and to take a fresh look at the radial
distribution and luminosity function of pulsars using these new PMPS
data, with the kind permission of Prof. R.N. Manchester, from
the ATNF Pulsar Catalogue of 1412 pulsars (Manchester et al. 2002). We
estimated the distances to the pulsars by using the NE2001 Galactic
electron density model (Cordes \& Lazio 2002, CL2002 hereafter).

In the space and luminosity distribution studies, it is very important
to make corrections for the observational selection effects. In this
contribution we used an empirical (and simple) method for the
correction of observational selection effects, as in Kodiara
(1974), Yusifov (1981), Leahy \& Wu~(1989), Wu \& Leahy (1989,
WL89 hereafter), with some modifications.

\section{Available Data and the Selection Effects}

At the time of preparing this contribution, the number of existing
pulsars was 1412. The ATNF Pulsar Catalogue contains the data of PMPS,
the Swinburne survey (Edwards et al. 2001) and all previous pulsar
survey results, and it supplies a good sample of data for the
statistical study of pulsars. Nearly 600 of them were discovered in the
Parkes and Swinburne Multibeam Pulsar Surveys (MBPS).

In this study we are interested mainly in statistics of ``normal"
pulsars. For this reason we excluded from our sample binary and
recycled ($\dot{P} <10^{-17}$ s/s), globular cluster and extragalactic
pulsars. In our study we mainly used pulsars in the regions of the
Galactic latitudes $|b|\leq 15^\circ$ and longitudes $-100^\circ\leq l
\leq 50^\circ$. In this study, as in the Galactic electron
distribution model NE2001, we used the IAU recommended value $R_\odot
= 8.5$~kpc as the distance Sun--GC.

The apparent distribution of pulsars for the subsequent corrections
due to selection effects is derived in the following manner. We drew
equidistant concentric circles around the Sun ($r_i$) and the GC
($R_j$) and made up a quasi-regular grid of points at the points of
intersection of these circles on the Galactic plane. Then we drew
circles of radius $R_{ci}$ around the grid points (or cells), counted
the number of pulsars within the boundaries of circle $R_{ci}$, and
estimated surface densities of pulsars around grid points. $R_{ci}$
increasing linearly is chosen from $0.1 R_\odot$ to $1/10$ of the
largest considered distances from the Sun (18.7 kpc) and calculated
by the relation $R_{ci}=0.85*(1.2*(i-1)/11+1)$, where $i$ varies from
1 to 12. Similar variation of cell radii naturally reduces the density
estimation errors at large distances due to distance uncertainties of
pulsars.

Making corrections for the selection effects, we used these data for
the calculation of the radial distribution of pulsars in the Galaxy.
Selection effects we divide into two categories and define them by the
relation:

\begin{equation}
\rho (r,R,l(r,R)) = K(l)K(r)\rho _o (r,R,l(r,R))\ ,
  \label{RorR}
\end{equation}
where $l$ is the Galactic longitude; $r$ and $R$ distances from the
Sun and the GC; $K(l)$ direction and $K(r)$ distance-dependent
selection factors; $\rho _o (r,R,l(r,R))$ is the true and $\rho
(r,R,l(r,R))$ is the observed distribution of surface densities of
pulsars on the Galactic plane.

\begin{figure}[b]
\begin{minipage}[t]{7.6cm}
\psfig{figure=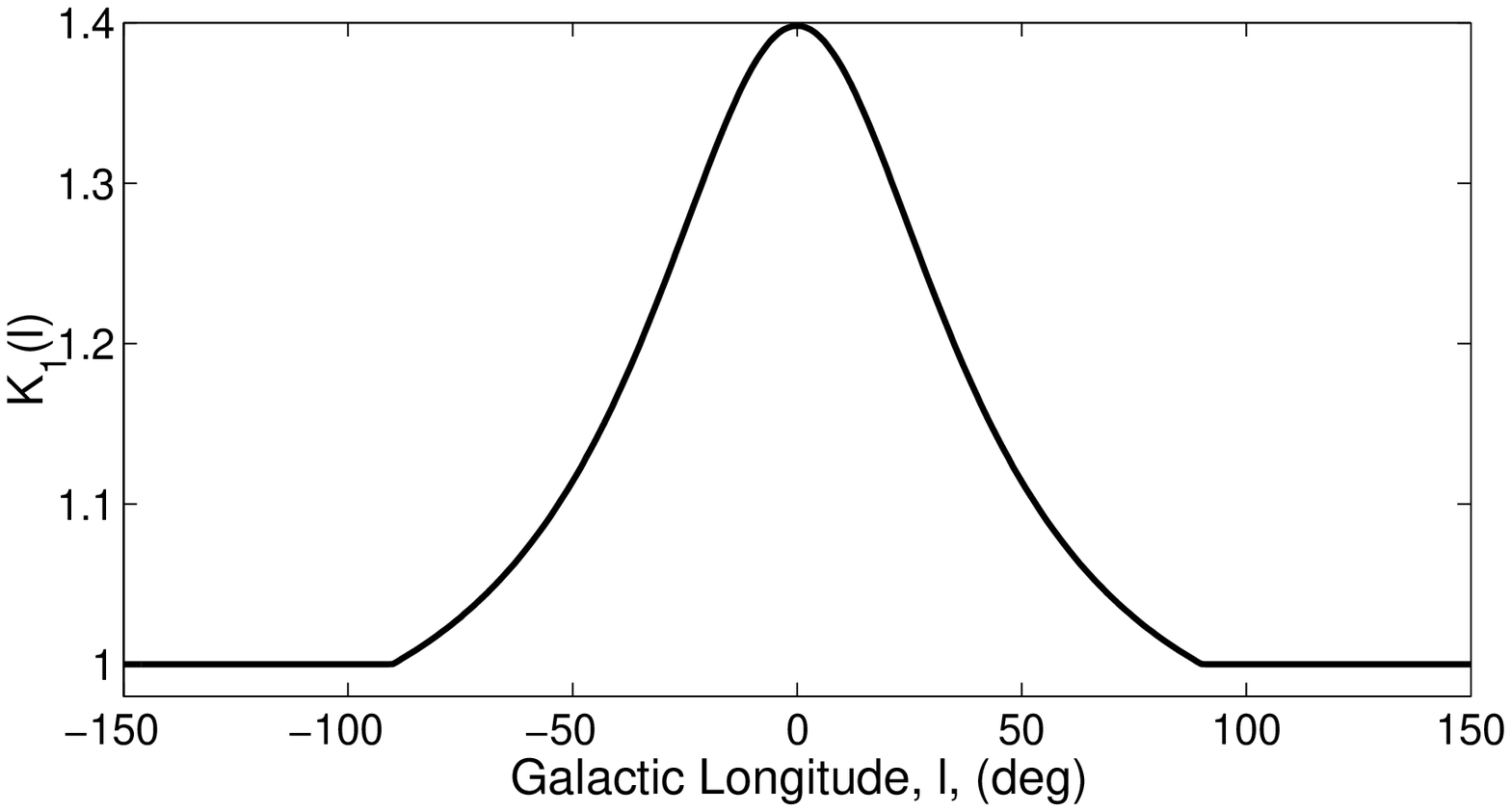,width=7.6truecm}
\caption{Dependence of the correcting factor $K_1( l )$
as a function of Galactic longitude.
\label{Fig:K1(l)}
}
\end{minipage}\hfill
\begin{minipage}[t]{7.6cm}
\psfig{figure=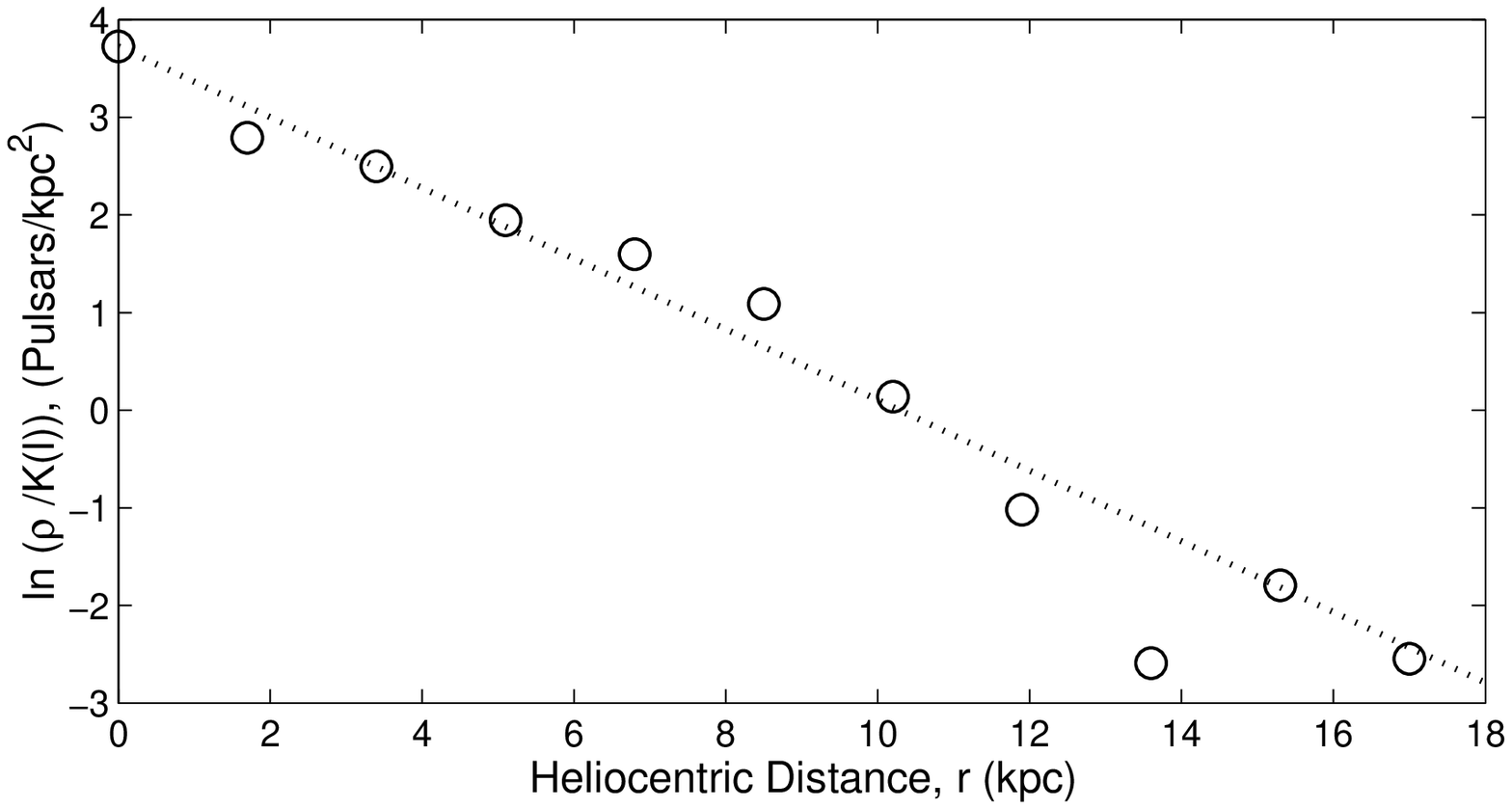,width=7.6truecm}
\caption{Variation of the surface densities of pulsars on the
galactocentric circle $R_\odot=8.5$ kpc at various distances from the Sun.
\label{Fig:K2(r)}
}
\end{minipage}
\end{figure}

$K(l)$ is connected with the background radiation which leads to
variations of the survey sensitivity with Galactic longitudes.
Neglecting latitude dependence as in WL89, the direction-dependent
correcting factor $K_1(l)$ is defined by the relation:

\begin{equation}
K_1(l)=1+ { T_{\rm sky\,1374}(l,b) \over T_{\rm R} }\ ,
\label{K1(l)}
\end{equation}
where $T_{\rm sky\,1374}(l,b)$ is the sky background temperature at
1374 MHz and $T_{\rm R} = 21$~K is the noise temperature of the
Parkes multibeam receiver (Manchester et al. 2001). The correction
factor $K_1( l )$ which was calculated from Eq.~(\ref{K1(l)}) is shown
in Fig.~\ref{Fig:K1(l)}. The observational number (or density) must be
corrected for this selection effect, multiplying by $K_1( l )$. It is
evident that $K(l)$ in Eq.~(\ref{RorR}) is equal to $K_1^{-1}(l)$. In
the anticenter direction and in the cell where the Sun is located,
$K_1( l )$ is assumed to be 1.

The distance-dependent selection factor $K(r)$ in
Eq.~(\ref{RorR}) is the function of pulse broadening caused by
scattering, scintillation etc. Here we make the simplifying assumption
that the combined effect of these factors may be described by the
exponential law as:

\begin{equation}
K(r) = K_1 (r)K_2 (r) \cdots K_i (r) = e^{ - c_1 r} e^{ - c_2 r}
\cdots e^{ - c_i r}  = e^{ - cr}\ ,
\label{K(r)} %\ref{K(r)}
\end{equation}
where $K_1(r)$, $K_2(r)$ etc. are the distance-dependent correction
factors relating to the pulse broadening, scattering and other
selection effects. The quantitative estimation of each of these
factors independently is difficult. But the combined effect of these
factors leading to reduced detection of pulsars away from the Sun may
be estimated empirically as described below.

Assuming that the surface density of pulsars is symmetric around the
GC, and considering a galactocentric circle with the radius
$R_\odot$, from Eq.~(\ref{RorR}) we obtain:

\begin{equation}
\rho (r,R_ \odot  ) = K(l)K(r)\rho _o (r,R_ \odot  )\ .
\label{RorRo} %\ref{RorRo}
\end{equation}
where  $\rho (r,R_\odot)$ is the observable density;
$K(l)$ is known and $\rho _o (r,R_ \odot )$ is constant.

In order to derive $K(r)$ from  Eq.~(\ref{RorRo}), in
Fig.~\ref{Fig:K2(r)}, we have plotted $\rho (r,R_\odot)/K(l)$ against
the heliocentric distance by circles.

Fitting the data in Fig.~\ref{Fig:K2(r)} by the least mean-squares
(LMS) method for the distance-dependent correction factor we obtain:

\begin{equation}
K(r)=\exp{(-cr)}\ ,
\label{K2(r)}
\end{equation}
where $c=0.362\pm 0.017$ and $ln\rho_o=3.73\pm0.16$.

\section{Radial Distribution and the Luminosity Function of Pulsars}

The true densities of pulsars were found from the apparent densities,
after correcting them by the relation (\ref{RorR}). $K(l)$ and $K(r)$
are calculated from Eqs.~(\ref{K1(l)}) and (\ref{K2(r)}) respectively.
The radial distributions of densities from the GC may be found by
averaging the densities $\rho (r,R)$ at various distances ($R_j$)
from the GC.

\begin{figure}[t]
\centerline{
\psfig{figure=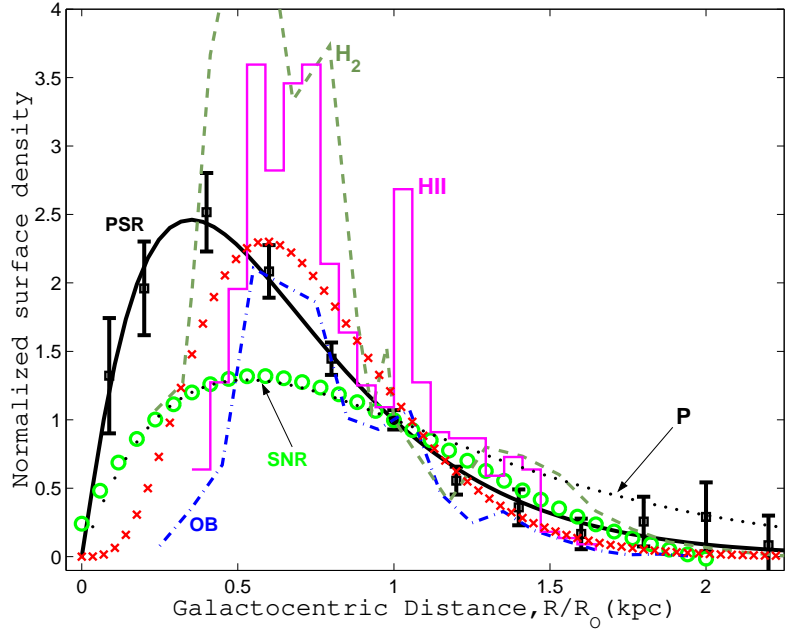,width=10.5truecm}
  }
\caption{Radial distributions of pulsars (squares) and other types of
Population~I objects: SNR distributions are from Case \& Bhattacharya
(1998); H2 column densities from Bronfman et al. (1988) and Wouterloot
et al. (1990); $\HII$ regions are from Paladini et al. (2004) (it must
be noted that $\HII$ regions represent the number of sources in the
0.5~kpc wide Galactocentric rings, but not surface densities); OB
star-formation regions data from Bronfman et al. (2000); marked by {\bf
P} are the radial distributions of birth location of NS (from Paczynski
1990); the expected radial distributions of birth location of NS
(Eq.~(\ref{Pach2})) are marked by crosses. }
\label{Fig:PopI}
\end{figure}

The results of obtained radial distribution of surface densities of
pulsars (squares) with corresponding error bars, together with the
distribution of other Population~I objects, are plotted in
Fig.~\ref{Fig:PopI}. In evaluating errors, we assumed that the number
of pulsars in cells, in a rough approximation, follows the Poisson
statistics. The errors for various data points are not equal,
so we calculated the weighted average of surface densities and
corresponding errors. In order to simplify the comparison with other
results, the densities in Fig.~\ref{Fig:PopI} are normalized to
the surface densities at the solar circle.

Within the galactocentric radius 0.5 kpc pulsars are absent; and from
the available data and precision of distance estimates, it is
difficult to estimate the density of pulsars there. But within the
range 0.5 kpc $<$ R $<$ 1 kpc there are 4 pulsars. Applying
the correction factors $K(l)$ and $K(r)$ to the apparent
density, we obtain $~50$ pulsars kpc$^{-2}$. In fitting the radial
distributions of pulsars, just this value will be used.

The radial distribution of pulsar surface densities in
Fig.~\ref{Fig:PopI} has been fitted by the  relation:

\begin{equation}
\rho(R)=A\biggl({R \over R_\odot} \biggr)^a
\exp{\biggl[-b\biggl({R-R_\odot \over R_\odot }\biggr) \biggr] }\ ,
\label{Gam14} %\ref{Gam14}
\end{equation}
where $R_\odot=8.5$ kpc is the Sun$-$GC distance. The best results of
the fits are: $ A=37.8\pm2.1~{\rm kpc}^{-2},\ a=1.12\pm0.10$ and
$b=3.20\pm0.24$.

Revising  Fig.~\ref{Fig:PopI}, it is seen that high frequency, high
sensitive PMPS searches revealed many pulsars around the GC, and
within error limits the derived density (50$\pm$16 pulsars kpc$^{-2}$)
is not less than that in the circumsolar region. Applying the Tauris
\& Manchester (1998, TM98 hereafter) beaming model, for this value we
obtain (500$\pm$150) pulsars kpc$^{-2}$.

In Fig.~\ref{Fig:PopI} we compare radial distributions of pulsars
and SNR with the radial distributions of their progenitors. It is seen
that, although the maxima of Population~I objects coincide, the
maximum of pulsar distribution is shifted to the GC by nearly
1.5~kpc. The maxima of pulsar and SNR distributions nearly coincide.
But the radial scale length (RSL) of pulsar distributions ($\sim
4$~kpc) is nearly two times less than the SNR distributions.

In Fig.~\ref{Fig:PopI} we also plotted by dots the radial distribution
of the birth location of NS which was derived by Paczynski (1990).
Although the maximum of this distribution coincides with the maximum of
the SNR distribution, its shape considerably deviates from the
distribution of Population~I objects. If the progenitors of NS are
OB-type Population~I stars, then from qualitative considerations of
Fig.~\ref{Fig:PopI}, it seems that the radial distribution of the birth
location of NS must be located near the line shown by crosses in
Fig.~\ref{Fig:PopI} and may be described by the relation:

\begin{equation}
\rho(R)=A\biggl({R \over R_\odot} \biggr)^a
\exp{\biggl[-b\biggl({R \over R_\odot }\biggr) \biggr] }\ ,
\label{Pach2} %\ref{Pach2}
\end{equation}
where $a=4$, $b=6.8$ and $R_\odot=8.5$ kpc is the Sun--GC distance.
The constant A must be chosen from the calibration constraints
(particularly, in the current case A = 1050).

In order to construct LF, we selected the sample of pulsars within a
cylinder of radius 2~kpc from the Sun. For faint pulsars we applied
standard volume correcting methods, and LF is calculated as

\begin{equation}
\Phi(L)=\Phi_O(L){V_{\rm max} \over V[<d(L)]}\ ,
\label{LF22} %\ref{LF22}
\end{equation}
where $\Phi(L)$ is the true LF, $\Phi_O(L)$ is the apparent LF,
$V[<d(L)]$ is the volume of sphere with radius $d(L)$ from the Sun in
which pulsars with luminosities $L$ are still observable,  $V_{\rm
max}=2 \pi d^2(L)z_{\rm max}$ is the volume of cylinder of radius
$d(L)$ around the Sun and $z_{\rm max}$ is the approximate value of the
maximum height of pulsars from the Galactic plane. We accept that
$z_{\rm max}=2$ kpc, and pulsars outside these altitudes are not more
than 2--5\% of observable pulsars.

\begin{figure}[t]
\centerline{\psfig{figure=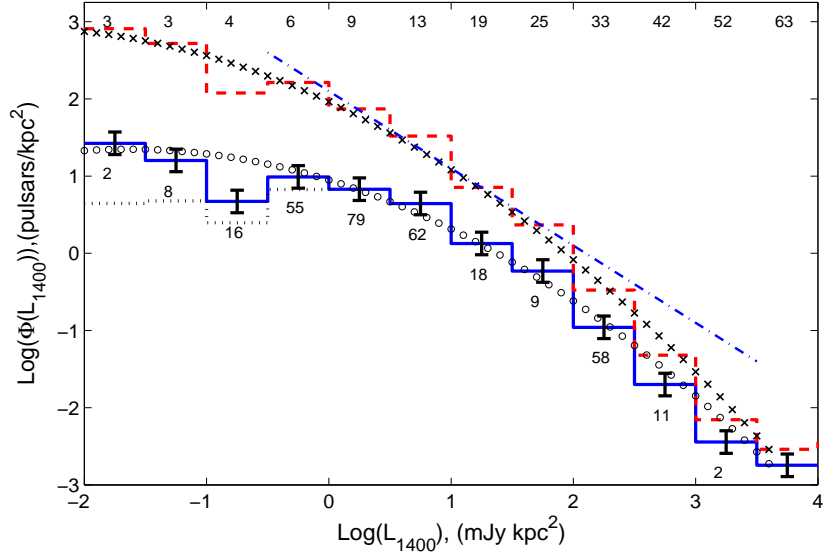,width=11truecm}}
\caption{Observed and corrected LF of pulsars. The observed
number of pulsars in the luminosity bins are indicated under the
corresponding bin. The observed distribution is given by the dotted
line. A solid line shows the volume-corrected distribution of
observable pulsars. The errors here represent the statistical
uncertainty (30\%) in distance estimates. A dashed line shows
the distribution of active pulsars, after applying beaming correction.
Lines marked by circles and crosses correspond to the Log--Normal
fitting of observed and corrected data by the relation (\ref{LF24}). A
dash-dotted line corresponds to the fitting of the data above
${\sim}1$ mJy kpc$^2$ assuming a slope of $-1$.}
\label{Fig:LF}
\end{figure}

The LF calculated on the basis of these data is presented in
Fig.~\ref{Fig:LF}. The total number of observable pulsars in every
luminosity bin is estimated on the basis of  all available pulsars,
but taking into account that high sensitive MBPS are carried out only
in the region of  $|b|\leq 15^\circ$ and $-100^\circ\leq l \leq
50^\circ$.

For the low-luminosity pulsars the LF is estimated from relation
(\ref{LF22}), where $z_{\rm max}$ is assumed to be 2 kpc. There are
only two pulsars with luminosities $>$ 100 mJy kpc$^2$ in the
considered region of d $<$ 2 kpc around the Sun. $\Phi(L)$ values for
them are calculated on the basis of large-distance pulsars revealed in
the MBPS, assuming that they are observable almost everywhere in
the Galaxy.

In this study we applied the TM98 beaming model for the correction
of LF. We estimated approximate values of the beaming fraction (BF)
corresponding to each luminosity bin and show them at the top of
Fig.~\ref{Fig:LF} in percent. LF, corrected according to these BFs, is
shown by the dashed line in Fig.~\ref{Fig:LF}.

The MBPS nearly doubled the number of pulsars around 0.1 mJy kpc$^2$,
but it is not enough to describe the LF by one straight line. As is
seen in Fig.~\ref{Fig:LF}, only the LF in limited regions of
luminosities may be fitted by a straight line. A flattening of LF at
low luminosities was already noted in previous studies and explained as
a result of the deficit of low-luminosity stars (see Lyne et al. 1985;
Guseinov \& Yusifov 1986; Cordes \& Chernoff 1997; Lyne et al. 1998,
and references therein). We fitted the data in Fig.~\ref{Fig:LF} by the
Log--Normal distribution before and after beaming corrections by one
analytical relation:

\begin{equation}
\Phi(L)={A_L \over \sigma_L \sqrt{2\pi}}
\exp{\biggl[-{1\over2}\biggl({{\rm log}L-{\rm log}L_O \over
\sigma_L}\biggr)^2 \biggr]}\ {\rm pulsars \; kpc}^{-2}\ .
\label{LF24} %\ref{LF24}
\end{equation}

For the constants of distributions $A_L, \sigma_L$ and ${\rm log}L_O$
before and after beaming corrections, we obtained values 67$\pm$15;
1.21$\pm$0.07;  $-$1.64$\pm$0.33  and  2600$\pm$600; 1.24$\pm$0.10;
$-$2.60$\pm$0.57 correspondingly.

\section{Results and Discussion}

We have studied the population of normal pulsars in the Galaxy,
using the ATNF Pulsar Catalogue where the distances are calculated
according to the new electron densities model NE2001.

We have derived the radial distribution of surface density of normal
pulsars by using the new distances. The maximum of radial distribution
is located at ${\sim}3.5$~kpc and ${\sim}1.5$~kpc nearer to the GC
relative to the maximum of distributions of Population~I objects.
Although the maximum of the distribution nearly coincides with
the maximum of distributions of their progenitors (SNRs), the RSL of
pulsars is  ${\sim}4$~kpc, i.e. nearly two times less than the SNR
distribution. Integrating the radial distribution for the total number
of normal pulsars with luminosities $L_{1400}\geq 0.1$ mJy
kpc$^2$ in the Galaxy, we obtained $(24\pm3)\times10^3$ and
$(240\pm30)\times10^3$ before and after applying beaming corrections.

The surface density of pulsars around the GC region is (50$\pm$16)
pulsars kpc$^{-2}$ and (500$\pm$150) pulsars kpc$^{-2}$ before and
after applying beaming corrections.

For the luminosities at 1400 MHz, the LF of pulsars is constructed
on the basis of pulsars within 2~kpc from the Sun. In estimating beaming
corrections we applied pulsar beaming models of TM98. The low
luminosity end flattening of LF became increasingly more evident. The
LF of pulsars  may be better described by the Log--Normal distribution
(Eq.~\ref{LF24}). Above a luminosity value of 0.1~mJy~kpc$^2$ for the
local surface density of normal pulsars we obtained $(41\pm5)$ pulsars
kpc$^{-2}$ and $(520\pm150)$ pulsars kpc$^{-2}$ before and after
beaming corrections. Assuming $\tau=10^7$yr as the mean life time of
active pulsars, for the corresponding birth-rate of pulsars in the
solar neighborhood we obtain 4 and 52 pulsars kpc$^{-2}$ Myr$^{-1}$
respectively before and after beaming corrections.

We recommend a new relation (Eq.~(\ref{Pach2})) for the expected
radial distribution of birth location of NSs, which is more closely
related to the radial distribution of Population~I objects.

The total number and frequency of generation normal pulsars obtained
from these data are nearly 1.5 and 4 times higher than previous
estimates, correspondingly before and after beaming corrections. One of
the reasons for the higher values is possibly connected with the
overestimated electron densities of the ISM. This leads to the necessity
of further improvement of the Galactic electron density model, which
requires $\HI$ measurements for distant pulsars.

One of the recent, but unsuccessful, attempts on pulsar distances and
LF estimations are the publications of Guseinov et al. (2002a, 2002b).
However, due to the fallacy of their method and approaches to the
problem and other numerous errors, their results are not discussed here.

A more complete report on this work will be given elsewhere
(Yusifov and K\"u\c c\"uk, in preparation).

\section*{Acknowledgments}

We would like to thank R.N. Manchester and the Parkes Multibeam
Pulsar Survey team for making the parameters of new pulsars available
on the internet prior to formal publication. We thank R. Wielebinski,
J.L. Han and F.F. \"Ozeren for reading the manuscript and for useful
discussions. We thank  Victor B. Cohen for help in preparing the
manuscript. This work has been partially supported  by Erciyes
University R/D project No. 01$-$052$-$1, Turkey. Extensive use was made
of both the Los Alamos preprint archive and the ADS system.

\section*{References}\noindent

\references

Bronfman, L., Cohen, R.S., Ajvarez, H., May, J., \& Taddeus, P.
  \Journal{1988}{\ApJ}{324}{248}.

Bronfman, L., Casassus, S., May, J., \& Nyman, L.-\AA\
  \Journal{2000}{\AAp}{358}{521}.

Case, J.L. \& Bhattacharya, D. \Journal{1998}{\ApJ}{504}{761}.

Cordes, J.M. \& Chernoff, D.F. \Journal{1997}{\ApJ}{482}{971}.
  %preprint [astro-ph/9706162].

Cordes, J.M. \& Lazio, T.J.W. (2002) preprint [astro-ph/0207156].

Edwards, R.T., Bailes, M., van Straten, W., \& Britton, M.C.
  \Journal{2001}{\ApJ}{326}{358}.

Guseinov, O.H., Yazgan, E., \"Ozkan, S., \& Tagiyeva, S.
  \Journal{2002a}{\em Rev. Mex. Astron. Astrof.}{39}{267}.
   %[astro-ph/0206030].

Guseinov, O.H., Yazgan, E., \"Ozkan, S., Tagiyeva, S., \& Yoldas, A.K.
  (2002b) preprint [astro-ph/0207306].

Guseinov, O.H. \& Yusifov, \,I.M. \Journal{1986}{\SA}{30}{47}.

Kodiara, K. \Journal{1974}{\PASJ}{26}{255}.

Kramer, M., Bell, J.F., Manchester, R.N., et al.
   \Journal{2003}{\MNRAS}{342}{1299}.
   %[astro-ph/0303473]

Leahy, D.A. \& Wu, X. \Journal{1989}{\PASP}{101}{607}.

Lyne, A.G., Manchester, R.N. \& Taylor, J.H. \Journal{1985}{\MNRAS}{213}{613}.

Lyne, A.G. , Manchester, R.N., Lorimer, D.R., Bailes, M., D'Amico, N.,
  Tauris, T.M., Johnston, S., Bell, J.F., \& Nicastro, L.
  \Journal{1998}{\MNRAS}{295}{743}.

Manchester, R.N., Lyne, A.G., Camilo, F., et al.
  \Journal{2001}{\MNRAS}{328}{17}.

Manchester, R.N., et al. (2002) {\em ATNF Pulsar Catalogue},
          http://www.atnf.csiro.au/research/pulsar/psrcat

Morris, D.J., Hobbs, G., Lyne, A.G., et al.
  \Journal{2002}{\MNRAS}{335}{275}.

Paczynski, B. \Journal{1990}{\ApJ}{348}{485}.

Paladini, R., Davies, R., \& DeZotti, G. (2004) in {\em From
Observations to Self-Consistent Modeling of the Interstellar Medium},
eds. M. Avillez \& D. Breitschwerdt, {\em Astrophys. \& Space Sci.}, in
press [astro-ph/0212341].

Tauris, T.M. \& Manchester, R.N. \Journal{1998}{\MNRAS}{298}{625}.

Wouterloot, J.G.A., Brand, J., Burton, W.B. \& Kwee, K.K.
  \Journal{1990}{\AAp}{230}{21}.

Wu, X. \& Leahy, D.A. (1989) {\em Acta Ap. Sinica} {\bf 9}, 233.

Yusifov, I.M. (1981) {\em Former Soviet Astr. Circ.} {\bf 1164}, 1.

\end{document}